\documentclass[A4]{aa}
\usepackage{graphicx}
\usepackage{times}

\begin{document}

\title{Simultaneous X-ray and optical observations of S5~0716+714 after the outburst of 
March 2004\thanks{Based on observations obtained with \emph{XMM-Newton}, an ESA science mission 
with instruments and contributions directly funded by ESA Member States and the USA (NASA).}}

\authorrunning{L. Foschini et al.} 

\author{L. Foschini\inst{1}, G. Tagliaferri\inst{2}, E. Pian\inst{3}, G. Ghisellini\inst{2}, 
A. Treves\inst{4}, L. Maraschi\inst{2}, F. Tavecchio\inst{2}, G. Di Cocco\inst{1}, S.R. Rosen\inst{5}}

\institute{INAF/IASF-Bologna, Via Gobetti 101, 40129 Bologna (Italy)
\and
INAF, Osservatorio Astronomico di Brera, Via Brera 28, 20121 Milano (Italy)
\and
INAF, Osservatorio Astronomico di Trieste, Via Tiepolo 11, 34131, Trieste (Italy) 
\and
Dipartimento di Scienze, Universit\`a degli Studi dell'Insubria, Via Valleggio 11, 22100, Como (Italy)
\and
Mullard Space Science Laboratory, University College of London Holmbury St Mary, Dorking, Surrey, RH5 6NT,  UK
}

\offprints{\email{foschini@iasfbo.inaf.it}}   
\date{Received 3 February 2006; Accepted 26 April 2006}

\abstract{At the end of March 2004, the blazar S5~$0716+714$ underwent an optical outburst that prompted for 
quasi-simultaneous target-of-opportunity observations with the \emph{INTEGRAL} and \emph{XMM-Newton} satellites.
In this paper, we report the results of the \emph{XMM-Newton} and \emph{INTEGRAL} OMC data analysis. The X-ray 
spectrum is well-represented by a concave broken power-law model, with the break at about $2$~keV. In the framework 
of the synchrotron self-Compton model, the softer part of the spectrum, which is described by a power law of index 
$\alpha \simeq 1.8$ ($f_\nu \propto \nu^{-\alpha}$), is probably due to synchrotron emission, while the harder part 
of the spectrum, which has $\alpha \simeq 1$, is due to inverse Compton emission. 
The blazar shows the long and short-term variability typical of low-frequency peaked BL Lac (LBL): the former is 
manifested by a gradual decrease in the optical flux from the peak as observed by ground
telescopes at the end of March 2004, while the latter is characterized by soft X-ray and optical flares on time 
scales from a few thousand seconds to few hours. We can follow spectral variations on sub-hour time scales and study 
their correlation with the flux variability.  
We find evidence that the peak energy of the time-resolved spectra is increasing with flux. The modeling 
of the spectral energy distribution compared with archival observations suggests that the long-term 
variability (from outburst to quiescence or viceversa) could be due to a change in the injected power, 
while the short-term variability (flares) could be explained with changes in the slope of the distribution 
of the electrons. 
}

\maketitle 

\section{Introduction}
The BL Lac object S5~$0716+714$ is one of the most variable blazars with intraday variability at radio, 
optical, and X-ray wavelengths suggesting a physical origin intrinsic to the source (e.g. Wagner \& Witzel 1995, 
Wagner et al. 1996, Montagni et al. 2006, Stalin et al. 2006). The optical continuum is so featureless that every 
attempt to determine the redshift has failed to date. The non-detection of the host galaxy first set a lower limit 
of $z > 0.3$ (Wagner et al. 1996) and more recently one of $z>0.52$ (Sbarufatti et al. 2005). 

The variability of this blazar is strong in every energy band: the optical and radio historical behaviour has been 
recently summarised by Raiteri et al.~(2003), while at the other end of the electromagnetic spectrum, the EGRET 
telescope onboard the \emph{Compton Gamma-Ray Observatory} (Hartman et al. 1999) detected several times S5~$0716+714$. 
Interestingly the strongest $\gamma-$ray detection, with a peak flux of 
$3.2\pm 0.7\times 10^{-7}$~ph~cm$^{-2}$~s$^{-1}$ (to be compared with the average EGRET value of 
$1.8\pm 0.2\times 10^{-7}$~ph~cm$^{-2}$~s$^{-1}$), was recorded in February 1995, the same year as a major 
optical outburst (with a peak around $R=12.8$) and small radio variations (cf Raiteri et al. 2003; see also 
Ghisellini et al. 1997). 

The X-ray observations have shown strong variations with short flares ($\approx 1000$~s) detected with \emph{ROSAT}
(Cappi et al. 1994). Broad-band observations with \emph{BeppoSAX} were performed in $1996$ and $1998$ during low 
optical activity (Giommi et al. 1999) and in $2000$ (Tagliaferri et al. 2003), triggered by a strong optical outburst, 
when the blazar displayed an increase of $2.3$ R-band magnitudes in $9$ days (cf. Raiteri et al. 2003). 
In the first two observations, S5~$0716+714$ was at a low flux state in the $0.1-10$~keV 
energy band ($3.4\times 10^{-12}$~erg~cm$^{-2}$~s$^{-1}$ in $1996$ and $4.4\times 10^{-12}$~erg~cm$^{-2}$~s$^{-1}$ in
$1998$) and was not detected at energies above $10$~keV. The soft X-rays (below $\approx 2.3$~keV) displayed significant
variability in agreement with the optical lightcurves, with the possibility that the optical and X-ray 
variations were not correlated, but the poor statistics prevented the authors from making firm conclusions 
(Giommi et al. 1999). 

\begin{figure}[!ht]
\centering
\includegraphics[scale=0.35,angle=270]{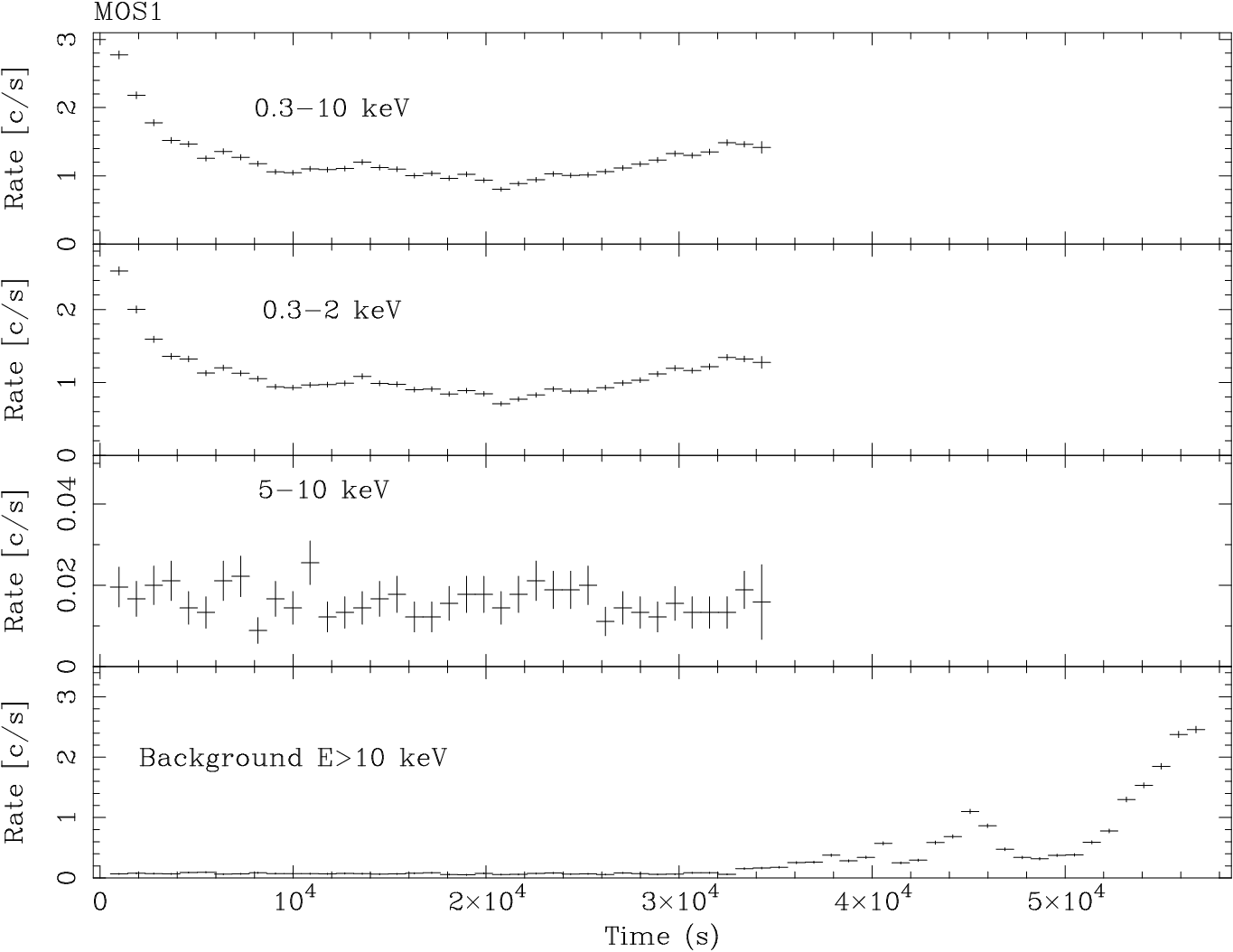}
\includegraphics[scale=0.35,angle=270]{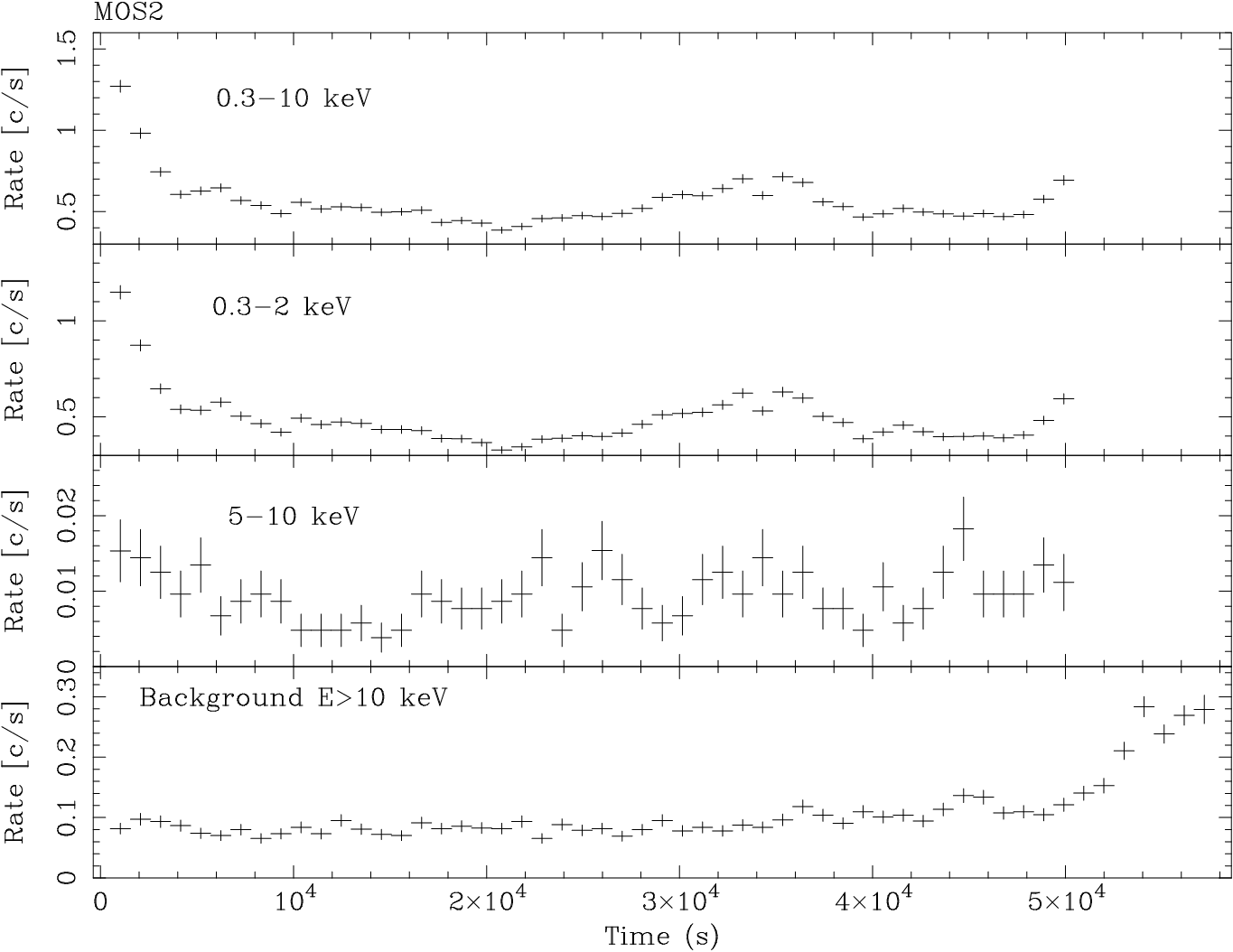}
\includegraphics[scale=0.35,angle=270]{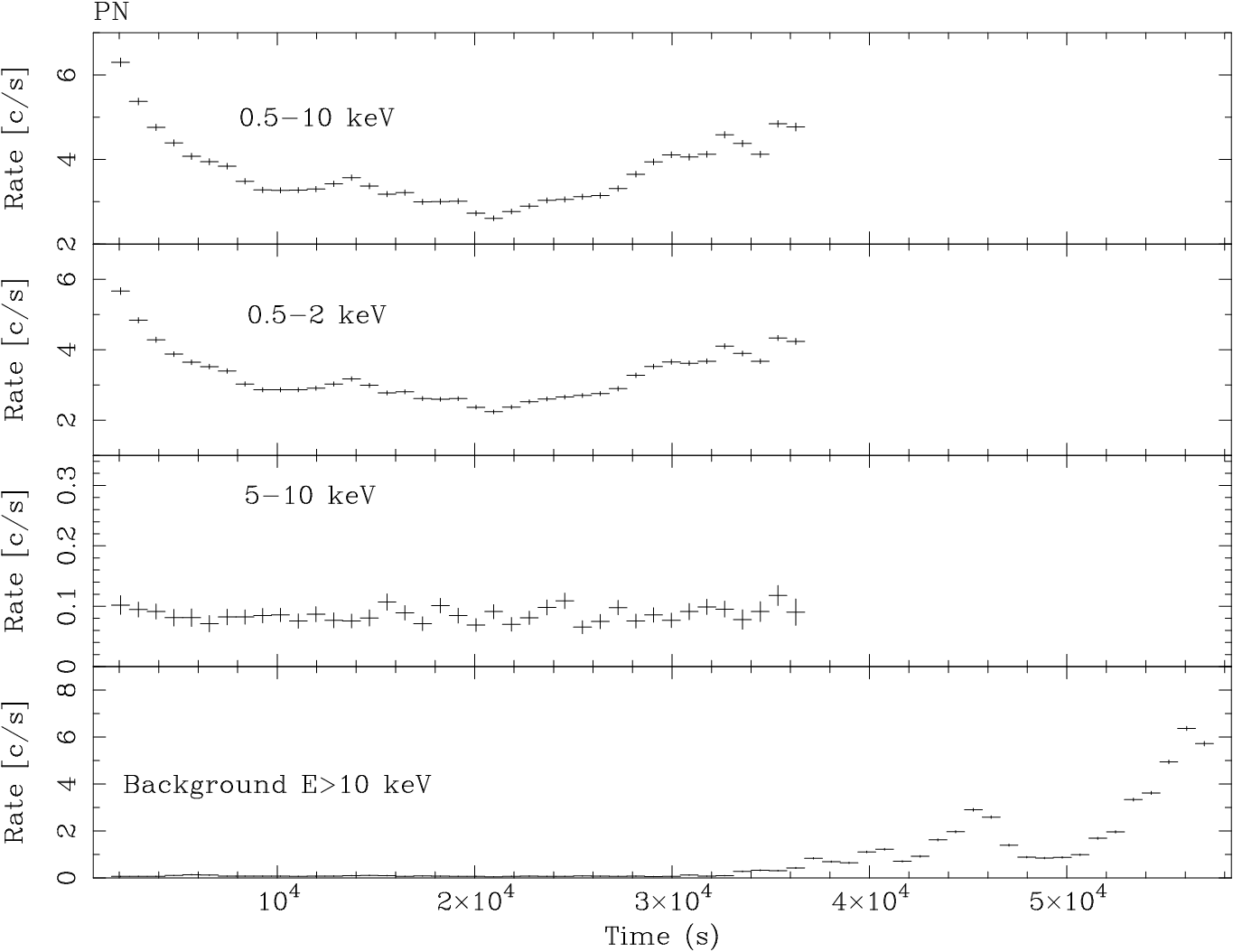}
\caption{EPIC MOS1 (\emph{top panel}), MOS2 (\emph{centre panel}), and PN (\emph{bottom panel}) lightcurves. 
For the two MOS detectors, the four panels indicate: source lightcurve in the energy bands $0.3-10$~keV, 
$0.3-2$~keV, and $5-10$~keV, plus the background lightcurve obtained from the whole detector count rate in the
energy band $E>10$~keV. For the PN detector, the four panels are: source lightcurve in the energy bands $0.5-10$~keV, 
$0.5-2$~keV, and $5-10$~keV, plus the background lightcurve obtained from the whole detector count rate in the
energy band $E>10$~keV. The time start is $4$ April $2004$ at $11^{\rm h}:10^{\rm m}:00^{\rm s}$ UTC. 
Time bins are $1040$~s wide for the MOS2 and $900$~s for MOS1 and PN.}
\label{epic_lc}
\end{figure}

The \emph{BeppoSAX} observation of 2000 confirmed the findings by Giommi et al. (1999), according to which the soft 
X-ray spectrum becomes steeper when the source flux increases. The source was also detected in hard X-rays 
up to $60$~keV. Again, variability appears to be present only in the soft X-rays (Tagliaferri et al. 2003).

In April $2004$, \emph{XMM-Newton} and \emph{INTEGRAL} observed S5~$0716+714$ nearly simultaneously, triggered by 
an optical outburst that occurred at the end of March.  The source was detected in the hard X-ray energy band 
(Pian et al. 2005). In Sect.~2 we report analysis of the \emph{XMM-Newton} EPIC camera data and describe 
the temporal and spectral results; in Sect.~3 we describe the results of the optical observations performed with 
the \emph{XMM-Newton} Optical Monitor and with the \emph{INTEGRAL} Optical Monitor Camera. The latter data were not 
presented in Pian et al. (2005). In Sect.~4 we describe the multiwavelength energy distribution of the blazar. The 
results are discussed in Sect.~5.

\section{X-ray data}
The BL Lac object S5~$0716+714$ was observed from $11:07$ UT of 4 April 2004 to $02:52$ UT of 5 April 2004 
(ObsID 0150495601, PI Tagliaferri) as part of a target-of-opportunity observation coordinated with \emph{INTEGRAL} 
(Pian et al. 2005). The EPIC MOS1 and MOS2 detectors (Turner et al. 2001) were set in small-window (with thin filter) 
and full-frame (with thick filter) mode, respectively. The EPIC PN detector (Str\"uder et al. 2001) was set in timing 
mode (thin filter), without imaging, but very high time resolution ($0.03$~ms) in order to avoid pile-up effects. 
For the processing, screening, and analysis of the data from the EPIC camera, standard tools were used 
(\texttt{XMM SAS v. 6.5}, with the latest calibration files, including the patch of August~$2005$, and HEAsoft 
\texttt{Xspec 11.3.2l} and \texttt{Xronos 5.21}). To analyse MOS data, we followed the standard procedures described 
in Snowden et al.~(2004), while we adopted the procedure outlined in Brinkmann et al. (2005) for PN (set in timing mode).

The nominal total exposure time was about $60$~ks; however, the high background reduced the effective exposure of MOS1 and 
PN to about $30$~ks. The MOS2 was relatively unaffected by the high background thanks to the use of the thick filter 
(Fig.~\ref{epic_lc}). To determine the start of high background periods precisely, we integrated the whole 
detector lightcurve at energies $E>10$~keV and excluded the periods with rates $>0.35$~c/s for MOS1 and $>1$~c/s for 
PN, as suggested by Kirsh (2005). For MOS2, we adopted a conservative threshold of $0.2$~c/s, which allowed us to obtain 
about $50$~ks of good exposure. 

\begin{table*}[!ht]
\caption{Spectral fits to the EPIC MOS1, MOS2, and PN data ($0.65-10$~keV) with a fixed absorption column 
($N_{H}=3.81\times 10^{20}$~cm$^{-2}$, Dickey \& Lockman 1990). The three time regions selected for time
resolved spectral analysis are shown in Fig.~\ref{mos_timereg}.
The uncertainties in the parameters correspond to a $90$\% confidence level for one parameter.}
\centering
\begin{tabular}{ccccccc}
\hline
Model($^\mathrm{a}$)   & $\Gamma_1$($^\mathrm{b}$)     & $\Gamma_2$($^\mathrm{c}$)  & $E_{\rm break}$($^\mathrm{d}$) & $\tilde{\chi}^2/dof$($^\mathrm{e}$) & $F_{\rm soft}$($^\mathrm{f}$) & $F_{\rm hard}$($^\mathrm{g}$)\\
\hline
\multicolumn{4}{c}{Whole observation (average)} & \multicolumn{3}{c}{MOS1+MOS2+PN}\\
\hline
PL      & $2.61\pm 0.01$ & {}                     & {}              & $1.91/1030$          & $1.1$          & $0.31$\\
BKPL    & $2.78\pm 0.02$ & $2.09_{-0.05}^{+0.04}$ & $2.0\pm 0.1$    & $1.08/1028$          & $1.2$          & $0.41$\\
\hline
\multicolumn{4}{c}{Period 1 ($0-2.5$~ks; high flux)} & \multicolumn{3}{c}{(MOS1+MOS2+PN)}\\
\hline
BKPL    & $2.86_{-0.04}^{+0.05}$ & $1.9_{-0.3}^{+0.4}$ & $2.9_{-0.8}^{+0.4}$ & $0.97/351$  & $2.0$          & $0.41$\\
\hline
\multicolumn{4}{c}{Period 2 ($20-25$~ks; low flux)} & \multicolumn{3}{c}{(MOS1+MOS2+PN)}\\
\hline
BKPL    & $2.75\pm 0.05$ & $1.93_{-0.08}^{+0.07}$ & $1.8_{-0.1}^{+0.2}$ & $1.14/545$       & $0.88$         & $0.38$\\
\hline
\multicolumn{4}{c}{Period 3 ($30-36$~ks; high flux)} & \multicolumn{3}{c}{(MOS2)}\\
\hline
BKPL    & $2.83\pm 0.09$ & $1.9_{-0.7}^{+0.3}$ & $2.6_{-0.4}^{+0.9}$ & $1.07/85$           & $1.4$          & $0.42$\\
\hline
\end{tabular}
\begin{list}{}{}
\item[$^{\mathrm{a}}$] \footnotesize{PL, simple power-law model; BKPL, broken power-law model.}
\item[$^{\mathrm{b}}$] \footnotesize{Photon index or soft photon index for PL and BKPL models, respectively ($f_E  \propto E^{-\Gamma}$).}
\item[$^{\mathrm{c}}$] \footnotesize{Hard photon index for BKPL.}
\item[$^{\mathrm{d}}$] \footnotesize{Break energy for BKPL [keV].}
\item[$^{\mathrm{e}}$] \footnotesize{Reduced $\chi^2$ and degrees of freedom.}
\item[$^{\mathrm{f}}$] \footnotesize{Unabsorbed flux in the $0.3-2$~keV energy band [$10^{-11}$~erg~cm$^{-2}$~s$^{-1}$].}
\item[$^{\mathrm{g}}$] \footnotesize{Unabsorbed flux in the $2-10$~keV energy band  [$10^{-11}$~erg~cm$^{-2}$~s$^{-1}$].}
\end{list}
\label{tab:spex}
\end{table*}

\subsection{Time variability}
We extracted the lightcurves from the three detectors of the EPIC camera. 
The time resolution of MOS1 (small window mode), MOS2 (full frame mode), and PN (timing mode) is $0.3$~s, $2.6$~s, and 
$3\times 10^{-5}$~s, respectively. 

We checked for pile-up effects with the \texttt{epatplot} task of the \texttt{XMM SAS} and found mild pile-up for the
MOS2 data (full frame mode), which could be removed by using an annular source region with inner and outer radii of $5''$ and 
$35''$, respectively. The MOS1 data (small window mode) are not affected by pile-up, so we adopted a circular region with 
$35''$ radius for the source spectra extraction. The background is extracted from a nearby source-free region of 
the same size. The source region was centred on the optical position of S5~$0716+714$, $\alpha=07:21:53.4$, 
$\delta=+71:20:36$ (J2000).

For the PN (set in timing mode and therefore without imaging capabilities, cf Brinkmann et al. 2005 for more details on the
analysis of this specific mode), we selected the columns $30\leq RAWX \leq 47$ for the source and $10\leq RAWX \leq 21$ 
for the background. For the latter, we used a smaller size to exclude a column containing a hot pixel. The background 
rates were then properly rescaled before subtracting them from the source-plus-background rates.

In order to minimize the impact of the background and other spurious effects, we selected only single pixel events 
($PATTERN==0$) tagged with $FLAG==0$, except for PN, since in timing mode it accepts only $PATTERN\leq 4$ 
(single and double pixel events). 

To investigate the X-ray variability, we built lightcurves in the energy bands $0.3-10$~keV, $0.3-2$~keV, 
and $5-10$~keV for MOS and $0.5-10$~keV, $0.5-2$~keV, and $5-10$~keV for PN. These values were selected 
according to the suggestion of the latest EPIC calibration report (Kirsch 2005) and to the source spectral 
characteristics (see next subsection). With a spectral break around $2-3$~keV, the selected energy 
bands are representative of the soft (synchrotron) and hard (inverse Compton) components of the source radiation. 
The soft X-ray flux shows strong variability (see Fig.\ref{epic_lc}): it initially decreases by a factor $2.5$ in 
a time scale of few hours, it undergoes a $\approx 10^4$~s duration flare of about $50$\%  amplitude, which is fully resolved 
only by the MOS2. At the end of the MOS2 light curve we note that the flux rises again. Some smaller amplitude and shorter 
time-scale variations are also observed.  A formal statistical analysis yields a Kolmogorov-Smirnov and $\chi^2$ probability 
of constancy approaching $0$ and a $RMS=27\pm 2$\% (MOS2 data, soft band lightcurve, with $2.6$~s bin, with negligible changes depending 
on the time bin). The hard X-ray emission ($5-10$~keV) does not show strong variability with $RMS\leq 20$\% (upper limit 
$3\sigma$).  

\begin{figure}[!t]
\centering
\includegraphics[scale=0.31,angle=270]{4959_f2a.ps}\\
\includegraphics[scale=0.31,angle=270]{4959_f2b.ps}
\caption{(\emph{upper panel}) EPIC MOS1, MOS2, and PN count spectra for the whole observation with 
residuals in terms of sigma (\emph{lower panel}). The solid curves represent the best-fit model (broken power law).}
\label{mos_spec}
\end{figure}

\begin{figure}[!t]
\centering
\includegraphics[scale=0.35,angle=270]{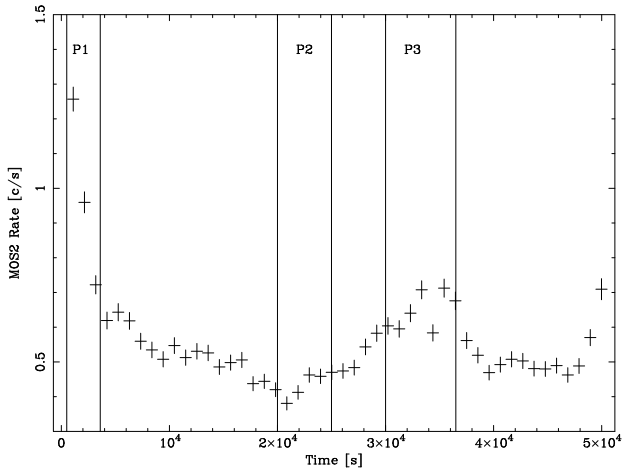}
\caption{EPIC MOS2 lightcurve ($0.3-10$~keV) with the time regions selected for the spectral analysis. The corresponding
spectral fits are shown in Table~\ref{tab:spex}.}
\label{mos_timereg}
\end{figure}

\begin{figure*}[!t]
\centering
\includegraphics[scale=0.7,angle=270]{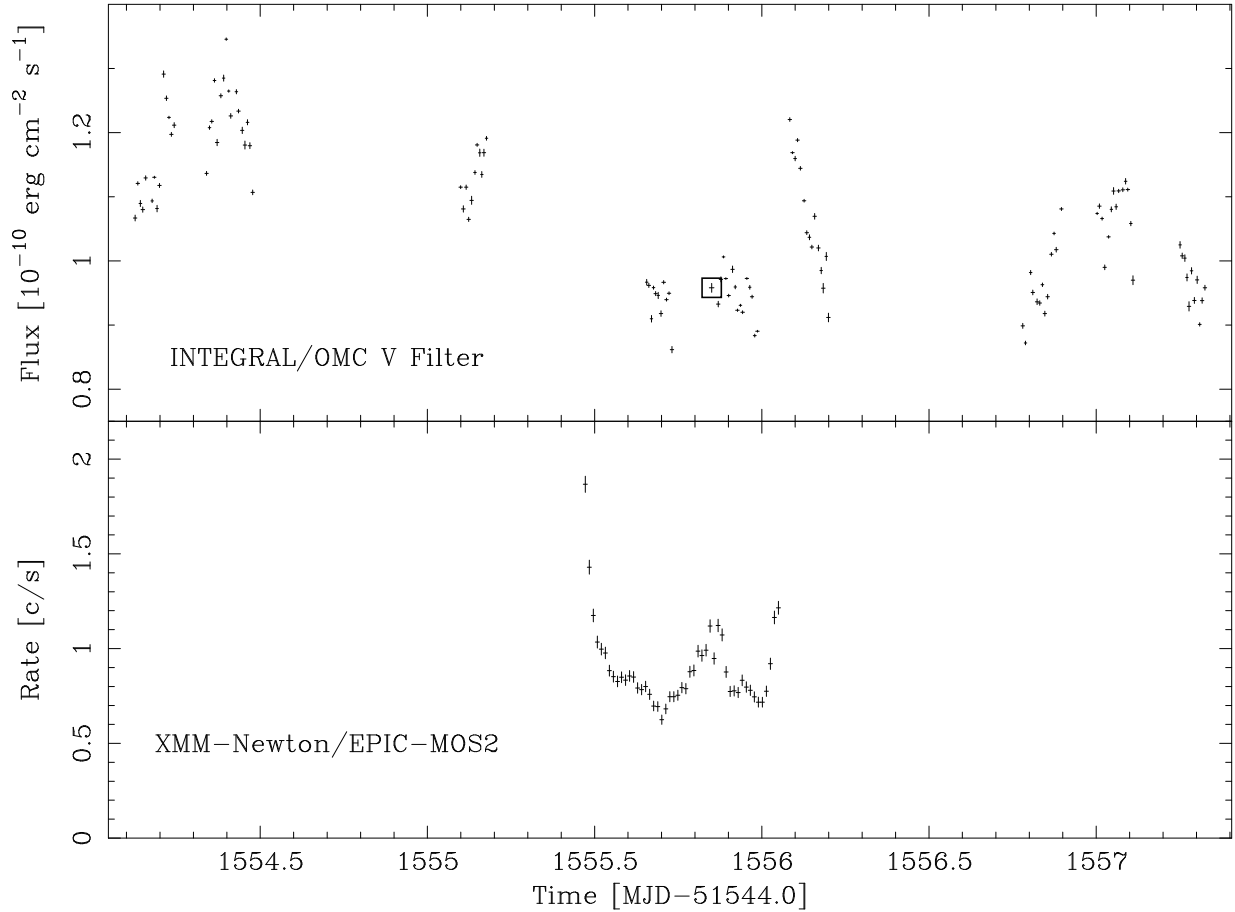}
\caption{(\emph{upper panel}) \emph{INTEGRAL}/OMC V filter lightcurve in the period $3-6$ April $2004$. The square
encloses the point from the OM onboard \emph{XMM-Newton} (see Table~\ref{tab:om}). (\emph{lower panel}) Simultaneous 
\emph{XMM-Newton} data from EPIC-MOS2 in the $0.3-10$~keV energy band.}
\label{fig:omc}
\end{figure*}

\subsection{Spectral characteristics}
The source counts for the spectral analysis were extracted from the same regions as were used for the lightcurves, while
the background counts were collected from a larger ($220''$) source-free region.
The spectra were rebinned so that each energy bin contained a minimum of 20 counts and fitted in the $0.65-10$~keV 
energy range (cf. Brinkmann et al. 2005 for PN in timing mode and Kirsch 2005 for MOS and intercalibration problems
between MOS and PN). The photon redistribution matrix and the related ancillary file were created appropriately with 
the \texttt{rmfgen} and \texttt{arfgen} tasks of \texttt{XMM SAS}.

The data from the three detectors were fitted together and the fit parameters are reported in Table~\ref{tab:spex}.
The single power-law model does not provide a good fit with evident residuals at high energy. The fit improves 
a lot with the broken power-law model, which we consider the best-fit model. Figure~\ref{mos_spec} shows the count 
spectra obtained with the data of the EPIC camera.

Since the lightcurves show significant variability on different time scales, we selected three time regions 
corresponding to high and low flux periods (see Fig.~\ref{mos_timereg}) and fitted the extracted
spectra with the broken power-law model (the fit with a single power-law is not acceptable for any
of the three spectra). The parameters of the three fits indicate the typical behaviour of LBL: 
as the flux rises, the spectrum at low energies becomes softer and the break energy increases, while
the hard-band spectral slope remains constant.

\section{Optical data}
Several exposures at different wavelengths were collected by the Optical Monitor (OM, Mason et al. 2001) onboard 
\emph{XMM-Newton} during the observation. We reprocessed the OM data with the SAS version also used for
the EPIC analysis, which gives the observed magnitudes already converted from OM count rates (the
region used for the photometry has a radius equal to $3$ times the FWHM of the PSF, which in turn is $\approx 2''$ for
UV and $\approx 1''.4$ for optical filters).
The results are summarised in Table~\ref{tab:om}. These measurements were used in the spectral
energy distribution (see the next section) and the magnitudes were dereddened by using
the extinction curve by Cardelli et al. (1989) with $A_V=0.102$ (Dickey \& Lockman 1990) and converted 
into flux by using the standard formulae (e.g. Zombeck 1990).

\begin{table}[!t]
\caption{Optical data of S5~$0716+714$ from the Optical Monitor (imaging mode) onboard \emph{XMM-Newton}.
All the exposures were about $1800$~s long. The uncertainties in the parameters are at $1\sigma$ level.}
\centering
\begin{tabular}{lcc}
\hline
Date/Time($^\mathrm{a}$) & Band($^\mathrm{b}$)   & magnitude($^\mathrm{c}$) \\
\hline
$04/04/2004$, $13:22$ & U      & $12.9\pm 0.1$\\
$04/04/2004$, $15:29$ & UVW1   & $12.8\pm 0.1$\\
$04/04/2004$, $18:06$ & UVM2   & $13.0\pm 0.1$\\
$04/04/2004$, $20:13$ & V      & $13.4\pm 0.1$\\
$04/04/2004$, $22:19$ & U      & $13.0\pm 0.1$\\
$05/04/2004$, $00:26$ & UVW1   & $13.0\pm 0.1$\\
$05/04/2004$, $02:00$ & UVM2   & $12.6\pm 0.1$\\
\hline
\end{tabular}
\begin{list}{}{}
\item[$^{\mathrm{a}}$] \footnotesize{Observation start date (dd/mm/yy) and time (UT).}
\item[$^{\mathrm{b}}$] \footnotesize{Band, with effective wavelength of $543$~nm for V, $344$~nm for U, $291$~nm for UVW1, and $231$~nm for UVM2.}
\item[$^{\mathrm{c}}$] \footnotesize{Observed magnitude (i.e. not dereddened).}
\end{list}
\label{tab:om}
\end{table}

\begin{figure*}[!t]
\centering
\includegraphics[scale=0.8]{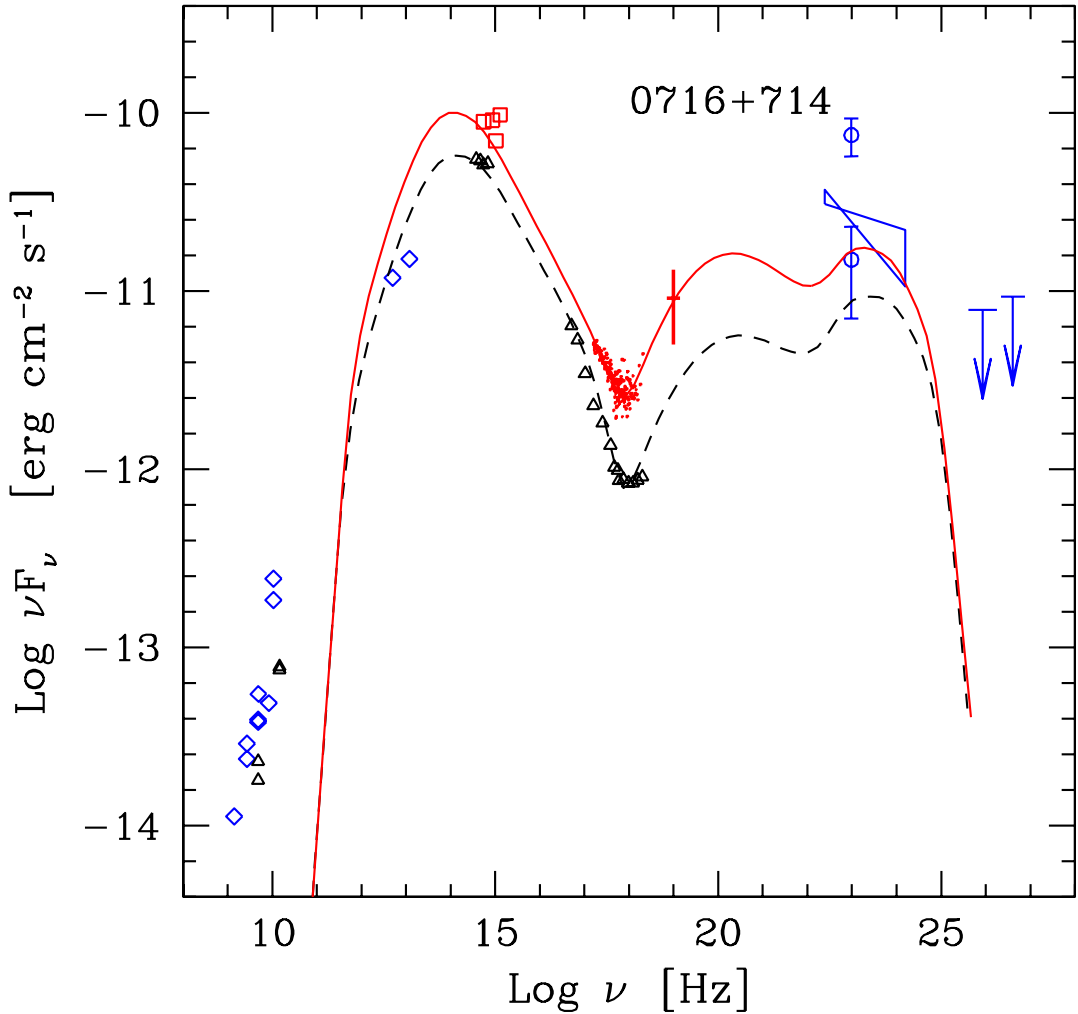}
\caption{Spectral energy distributions of S5~$0716+714$ at various epochs. 
The cross, points, and open squares represent simultaneous \emph{INTEGRAL}/ISGRI 
(Pian et al. 2005) and \emph{XMM-Newton} EPIC and OM (present work) observations in April 2004, respectively.
The triangles represent \emph{BeppoSAX}, optical, and radio nearly simultaneous observations in 1996 
(Giommi et al. 1999). Non simultaneous EGRET data are also reported (Hartman et al. 1999), referring to 
the minimum, maximum, and average flux (open circles). The latter also includes the photon index interval. Radio data 
(diamonds) and TeV upper limits are also non simultaneous (see Tagliaferri et al. 2003). 
The lines represent the model by Ghisellini et al. (2002) adapted to fit the April $2004$ 
(continuous line) and November 1996 (dashed line) energy distributions.}
\label{fig:sed}
\end{figure*}

The Optical Monitor Camera (OMC, Mas-Hesse et al. 2003) onboard the \emph{INTEGRAL} satellite 
provides optical data with a standard Johnson V filter simultaneously with the high-energy data 
of the on-axis source. The OMC PSF is $25''$-sized (FWHM), and we checked the USNO B1 Catalog (Monet et al. 2003)
for contaminating sources inside a region with twice this radius.  We found $4$ sources, 
in addition to the blazar, with $B$ magnitudes around $20$ exhibiting no significant variability 
between the 2 epochs at which the USNO survey was performed, so that their contribution can be neglected.

The V filter lightcurve of S5~$0716+714$ (Fig.~\ref{fig:omc}), obtained from the preprocessed data of the 
OMC public archive\footnote{\texttt{http://sdc.laeff.esa.es/omc/}. Based on data from the \emph{INTEGRAL}/OMC 
Archive at LAEFF, processed by ISDC. \emph{INTEGRAL} is an ESA mission with instruments and a science data centre 
funded by ESA member states (especially the PI countries: Denmark, France, Germany, Italy, Switzerland, Spain), 
the Czech Republic, and Poland, and with the participation of Russia and the USA.}, was dereddened with
$A_V=0.102$ (Dickey \& Lockman 1990) and converted into flux by using the standard formulae (e.g. Zombeck 1990).

\emph{INTEGRAL} observed the blazar twice between $10-17$ November $2003$ (ObsID $0120177$, PI Wagner) and between 
$2-7$ April $2004$ (ObsID $0220049$, PI Pian). During the November 2003 observation, S5~$0716+714$ was 
in an optically faint state ($R=14.17-13.64$) and there was no detection at hard X-rays (Ostorero et al. 2006, see also
Beckmann et al. 2006).

In the second \emph{INTEGRAL} observation (April 2004), the optical flux was higher, consistent with the one measured 
from the \emph{XMM-Newton} OM (see Fig.~\ref{fig:omc}) and with hard X-rays detection (Pian et al. 2005). A  general 
decrease over the observation duration is modulated by shorter time scale flares, with flux variations of the order 
of 30\% (Fig.~\ref{fig:omc}).

\begin{figure*}[!ht]
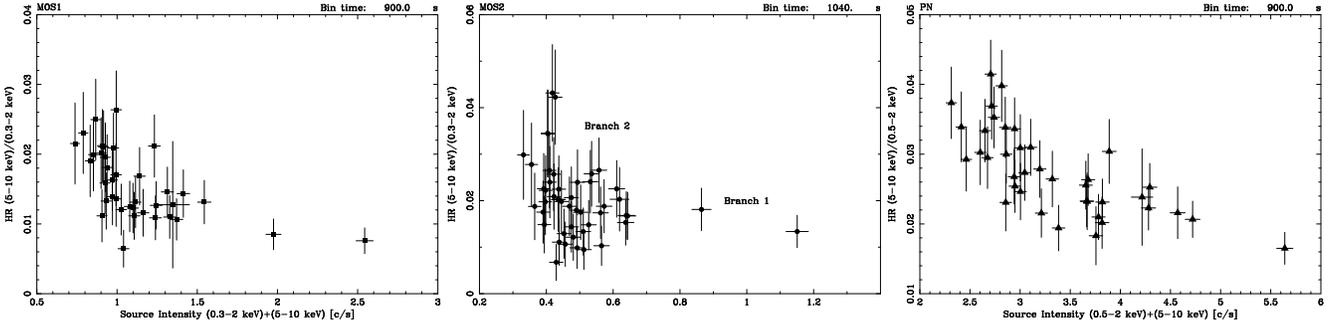

\centering
\includegraphics[scale=0.25,angle=270,clip,trim=0 0 20 0]{4959_f6a.ps}
\includegraphics[scale=0.25,angle=270,clip,trim=0 0 20 0]{4959_f6b.ps}
\includegraphics[scale=0.25,angle=270,clip,trim=0 0 20 0]{4959_f6c.ps}
\caption{Source intensity versus hardness ratio -- defined as $(5-10)/(0.3-2)$~keV -- for MOS1 (\emph{left panel}), MOS2 
(\emph{centre panel}), and for PN (\emph{right panel}), but in the energy bands $(0.5-2)$ and $(5-10)$~keV.}
\label{fig:hr}
\end{figure*}

\section{Spectral energy distribution}
In Fig.~\ref{fig:sed} we report the spectral energy distributions of S5~$0716+714$ constructed with historical data (see 
Tagliaferri et al. 2003) and with the \emph{INTEGRAL} and \emph{XMM-Newton} data acquired in April 2004. We adopted the 
model described in Ghisellini et al. (2002), to which we refer the reader for a detailed explanation of the parameters, and assumed 
a redshift $z=0.3$ for S5~$0716+714$, in order to compare the present results with similar modeling performed on the past 
observations (cf. Table~4 in Tagliaferri et al. 2003). 

The model envisages synchrotron radiation and inverse Compton scattering as responsible for the
spectral emission at low (radio to UV) and high (X- and gamma-rays) frequencies.  The geometry of the
emitting region is a cylinder with radius $R=2\times 10^{16}$~cm and height 
$\Delta R' \approx R/\Gamma=1.3\times 10^{15}$~cm. The bulk Lorentz factor is $\Gamma=15$
and the viewing angle $\theta = 3.4^{\circ}$, implying a Doppler factor $\delta=16.7$. The magnetic field is $3$~G and 
the electron distribution is assumed to be $N(\gamma)\propto \gamma^{-n}$. For the present case there are three 
different distribution regions:

\begin{itemize}
\item $N(\gamma)\propto \gamma^{-2}$, for $\gamma < \gamma_{\rm break}$;

\item $N(\gamma)\propto \gamma^{-2.9}$, for $500 = \gamma_{\rm break} < \gamma < \gamma_{\rm cool}$;

\item $N(\gamma)\propto \gamma^{-3.9}$, for $1424 = \gamma_{\rm cool} < \gamma < \gamma_{\rm max} = 3.7\times 10^4$;

\end{itemize}

\noindent where $\gamma_{\rm cool}$ is the Lorentz factor for which the radiative cooling time is equal to $\Delta R'/c$.
The injected power in the comoving frame is $L=4\times 10^{42}$~erg/s. 

The present parameters imply minimal changes with respect to the model parameters of the $1996$ observation
when the source was in low activity, except for the injected power, which in $1996$ was about half the present value 
($2.2\times 10^{42}$~erg/s), and for some changes in $\gamma_{\rm max}$. For the observation in $2000$, when the source was 
active, the model requires similar parameters and an injected power of $5\times 10^{42}$~erg/s, very similar to the 
present case. For more details on the model parameters of the previous \emph{BeppoSAX} observations see Tagliaferri et 
al. (2003).

\section{Discussion}
\emph{XMM-Newton} observed S5~$0716+714$ in April 2004 after an optical outburst and detected it in a high soft X-ray state.  
The X-ray flux in the $0.1-2$~keV band is  about a factor of 5 and a factor of 2 higher than detected by \emph{BeppoSAX} 
in $1996$ and $1998$, and in $2000$, respectively.  The $2-10$~keV flux appears stabler: it is comparable to the
$2000$ level and a factor 2 higher than in 1996 (Giommi et al. 1999; Tagliaferri et al. 2003).  The hour time-scale 
variability of the X-ray flux observed during the present observation confirms that the soft X-ray variability has 
larger amplitude than the hard X-ray variability (Fig.~\ref{epic_lc}).  This suggests that different emission components 
are responsible for the soft and hard X-rays, as already proposed by the previous authors.

Accordingly, the X-ray spectrum is not consistent with a single power-law, and is rather described by a broken power-law. 
The soft X-ray emission represents the higher energy tail of a synchrotron spectrum, and is therefore highly variable, 
while the hard X-ray emission, which has a flat spectrum, detected up to $60$~keV (Tagliaferri et al. 2003; Pian et al. 
2005), is less variable, as it originates from  inverse Compton scattering of relativistic electrons off modestly 
variable infrared and radio synchrotron radiation. The X-ray hour time scale spectral variations support this 
interpretation by providing evidence for the shift of the break energy toward higher energies during flares. In 
order to corroborate our short time scale spectral findings, we have also constructed intensity-hardness ratio plots 
(Fig.~\ref{fig:hr}). These display the typical characteristics of other well known Low-Energy Peaked BL Lacs (LBL), 
like BL Lac itself (Ravasio et al. 2002, 2003) and ON 231 (Tagliaferri et al. 2000). The trend is very close to the 
\emph{BeppoSAX} behaviour (cf Fig.~4 in Giommi et al. 1999), with high hardness at low fluxes.

The samplings of the \emph{XMM-Newton} EPIC and \emph{INTEGRAL} OMC lightcurves do not coincide exactly and,
particularly for the optical data, several time gaps are present, therefore it is difficult to establish a correlation 
between the two. The decreasing trend of the OMC lightcurve suggests that S5~$0716+714$ is returning to its quiescence 
state after the outburst at the end of March $2004$. The short flare at the end of the \emph{XMM-Newton} EPIC lightcurve 
seems to have a counterpart in the optical (Fig.\ref{fig:omc}), but the gap at the beginning of the optical 
flare does not allow us to evaluate the real start of the event.

We note two types of activity of S5~$0716+714$, labelled in Fig.~\ref{fig:hr} (\emph{centre panel}) 
with ``Branch 1'' and ``Branch 2'', respectively: the first type (Branch 1) is likely to be related to the activity 
following the major optical outburst that occurred at the end of March $2004$ (these points indeed refer to the very
beginning of the X-ray lightcurves shown in Fig.~\ref{epic_lc}). We already noted that, as shown in Fig.~\ref{fig:omc}, 
the optical flux measured by OMC is gradually declining. Moreover, Pian et al. (2005) report that the $30-60$~keV 
flux during the first $\approx 80$~ks of the \emph{INTEGRAL} observation is greater than the average over the whole 
obsevation. Therefore, although the position of the branch 1 in the intensity-hardness ratio plots favours the hypothesis
that can be the final tail of the outburst, we cannot exclude that we are seeing the decaying part of a major flare 
occurred during the gradual declining of the source after the outburst.

The second type of activity (Branch 2) is linked to the short flares randomly occurring during the long term
activity. Similar behaviour has been already observed in other 
BL Lac objects, like the high-energy peaked BL Lac (HBL) Mkn~$421$ (e.g. Sembay et al. 2002, Brinkmann et al. 2003, 
Ravasio et al. 2004, Brinkmann et al. 2005) or PKS~$2155-304$ (Zhang et al. 2006).
However, it is worth noting the different origins of this behaviour: in HBL, the X-ray emission is due to the synchrotron 
radiation, while in LBL there is competition between synchrotron and inverse Compton. The HBL display variability both 
in the soft and hard X-ray energy bands, with more significant variability at high energy: higher fluxes correspond to 
harder spectra, reflecting changes in the particle injection. On the other hand, the LBL are generally variable only in 
the soft X-rays, while the hard X-ray energy band shows little or no variability, so the hardness ratio 
reflects the dominance of synchtrotron relative to inverse Compton radiation. 

As known for other LBL, the soft X-rays' significant variability and the short flares in S5~$0716+714$
are due to the most energetic electrons of the synchrotron spectrum, cooling much faster than the electrons producing 
the inverse Compton emission, which are likely to be at the low end of the distribution. The gradual decay afterburst 
is instead likely to be attributed to the escape of electrons from the processing region or to a decrease in the soft 
seed photons or both. 

Instruments with higher senstivity, particularly in the hard X-ray and $\gamma-$ray domains, are necessary to follow 
the evolution of the source completely. The next generation of satellites operating in the $\gamma-$ray energy band, 
like GLAST\footnote{\texttt{http://www-glast.stanford.edu}}, and in the hard X-ray band, like SIMBOL-X (Ferrando et al. 2005), 
should provide the optimal temporal resolution for monitoring the short-term blazar variability, which appears to be 
the key to understand the mechanisms at work in this class of sources.

\begin{acknowledgements}
LF wishes to thank R. Gonzalez Riestra of \emph{XMM-Newton} SOC for discussion of EPIC background
and P. Giommi for discussion of previous \emph{BeppoSAX} observations. Thanks also to the
anonymous referee for helpful criticism that improved the manuscript.
This work was partially supported by the Italian Space Agency (ASI) under contract I/R/046/04. 
\end{acknowledgements}

\end{document}